\documentclass[prl,superscriptaddress,amsmath,amssymb,twocolumn]{revtex4-2}

\usepackage{lmodern}
\usepackage{graphicx}% Include figure files
\usepackage{mathtools}
\usepackage{amsmath}
\usepackage{amsfonts}
\usepackage{amssymb}
\usepackage{float}
\usepackage{cancel}
\usepackage{subfigure}
\usepackage{adjustbox}
\usepackage{dcolumn}% Align table columns on decimal point
\usepackage{bm}% bold math
\usepackage{hyperref}% add hypertext capabilities
\hypersetup{linktocpage,
    colorlinks=true,
    linkcolor=blue,
    citecolor=blue,
    urlcolor=blue,
    pdfdisplaydoctitle=true,
    pdfpagemode=UseOutlines,
    bookmarksnumbered=true}
\usepackage{mathrsfs,dsfont}
\usepackage{color}
\usepackage{wrapfig}
\usepackage{comment}
\begin{document}

\title{Earthquake-like dynamics in ultrathin magnetic film}

\author{Gianfranco Durin}
\affiliation{Istituto Nazionale di Ricerca Metrologica, Strada delle Cacce 91, Torino, 10135, Italy}%

\author{Vincenzo Maria Schimmenti}
\affiliation{LPTMS, CNRS, Universit\'e  Paris-Saclay,  91405,  Orsay, France}
\affiliation{Laboratoire Interdisciplinaire des Sciences du Num\'erique (LISN), TAU team,Gif-sur-Yvette,91190,France }%
\author{Marco Baiesi}
\affiliation{Department of Physics and Astronomy, Via Marzolo 8, Padova, 35131, Italy}
\affiliation{INFN, Sezione di Padova,Via Marzolo 8, Padova, 35131, Italy}%
\author{Arianna Casiraghi}
\affiliation{Istituto Nazionale di Ricerca Metrologica, Strada delle Cacce 91, Torino, 10135, Italy}%
\author{Alessandro Magni}
\affiliation{Istituto Nazionale di Ricerca Metrologica, Strada delle Cacce 91, Torino, 10135, Italy}%
\author{Liza Herrera-Diez}
\author{Dafin\'{e} Ravelosona}
\affiliation{Centre de Nanosciences et de Nanotechnologies,Universit\'e Paris-Saclay, CNRS,Palaiseau, 91120, France } %
\author{Laura Foini}
\affiliation{Institut de Physique Th\'eorique, Universit\'e Paris-Saclay, CNRS, Gif-sur-Yvette, 91191, France}%
\author{Alberto Rosso}
\affiliation{
LPTMS, CNRS, Universit\'e  Paris-Saclay,  91405,  Orsay, France
}

\begin{abstract}
We study the motion of a domain wall on an ultrathin magnetic film using the magneto-optical Kerr effect (MOKE). At tiny magnetic fields, the wall creeps only via thermal activation over the pinning centers present in the sample. Our results show that this creep dynamics is highly intermittent and correlated. A localized instability triggers a cascade, akin to aftershocks following a large earthquake, where the pinned wall undergoes large reorganizations in a compact active region for a few seconds.
Surprisingly, the size and shape of these reorganizations display the same scale-free statistics of the depinning avalanches in agreement with the quenched Kardar-Parisi-Zhang universality class.
\end{abstract}

\date{\today}

\maketitle
An important class of future spintronic nanoelectronic devices is based on fully controlling magnetic domain walls in ultrathin films \cite{PAR-15, GU-22}. When used as memory devices, for instance, it is fundamental to control their position stability and understand their dynamics under a small perturbation. It is well known that defects naturally present in the nanostructure can pin the domain wall. Consequently, the wall creeps at a finite temperature, with a velocity strongly vanishing with the applied magnetic field. 
In ultrathin magnetic films, the creep regime holds up to room temperature and well below the depinning  field $H_{\text{dep}}$. After an initial transient, the wall moves with a small, steady velocity given by the celebrated {\em creep formula}:
\begin{equation}
\label{eq:creep_formula}
\ln v(H) =  \left(\frac{H}{H_0} \right)^{-1/4} +\ln v_0
\end{equation}
Here, $H_0$  and $v_0$ are materials and temperature-dependent parameters. The exponent $1/4$ is instead universal and is the true hallmark of the creep dynamics. It has been first predicted in~\cite{IV-87}, measured in~\cite{LEM-98} over several decades of velocity, and then confirmed in many experiments \cite{KIM-09,JEU-16}. Despite this success, the nature of the creep dynamics remains controversial. In particular, several hypotheses are made on the length scales involved and on the shape of the wall.  

The original derivation of the creep formula assumes that the small magnetic field tilts the minima but leaves the landscape locally at thermal equilibrium. Within this picture, one finds a single length $L_{{\rm opt}}$, the scale of the reorganization needed to overcome the optimal energy barriers. Under this assumption, one can estimate that $L_{{\rm opt}} \sim H^{-3/4}$ and the corresponding energy barriers grow as $H^{-1/4}$  ~\cite{IV-87,DMMAV,AGO-12,FER-21}.
%The creep velocity (\ref{eq:creep_formula}) is thus determined by the activation time over such barriers $\sim  H^{-1/4}$ and depends solely on the property of the wall at thermal equilibrium. 
Below the scale $L_{{\rm opt}}$, the dynamics is thus purely thermal, characterized by an incoherent back-and-forth motion. Above $L_{{\rm opt}}$ instead, the wall never comes back and undergoes a novel slow reorganization of size $L_{{\rm opt}}$ in a different location~\cite{IV-87}. 
 
Further studies, based on functional renormalization group (FRG)~\cite{CHA-00} and numerical simulations at infinitesimal temperature $T\to 0^+$~\cite{KOL-09,FER-17, PUR-17, FER-21}, have proposed a different scenario. The activation over a size $L_{\rm opt}$ destabilizes the local energy landscape and reduces the size of the energy barriers. Similarly to what is observed in earthquakes, the jump of size $L_{\rm opt}$ acts as the mainshock that produces a cascade of aftershocks of smaller size \cite{BAI-04, JAG-10, JAG-14, SCH-19}. Hence,  the region undergoes a much larger reorganization and mimics the depinning avalanches belonging to the quenched Edwards-Wilkinson (qEW) universality class \cite{FIS-98,KAR-98}. The scale-free statistics of these avalanches is valid up to a length $L_{\text{av}}$ much more extensive than $L_{{\rm opt}}$ and controlled by the finite values of the temperature and the field.

Interestingly, this scenario has strong connections with the thermal facilitation proposed to justify the dynamical heterogeneity in glass-forming liquids \cite{OZA-23, TBOPW-PP-23}. The mechanism is similar: the slow relaxation time is dominated by localized slow events that nucleate large responses on a much larger scale. However, experimental evidence of these large reorganizations is still lacking.

\begin{figure*}[ht]
\textbf{(a)}
\includegraphics[height=5.5cm]{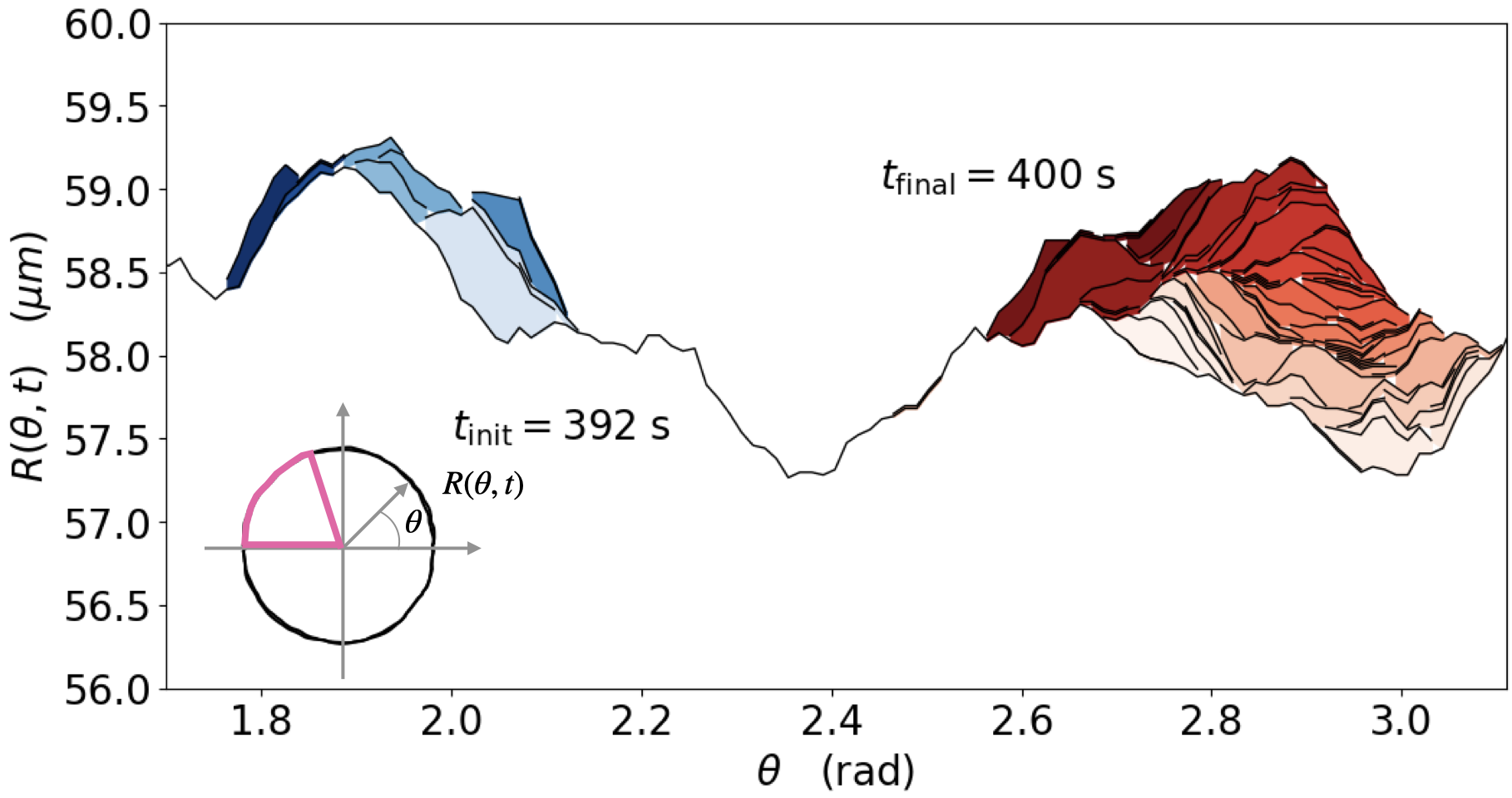}
\textbf{(b)}
\includegraphics[height=5.5cm]{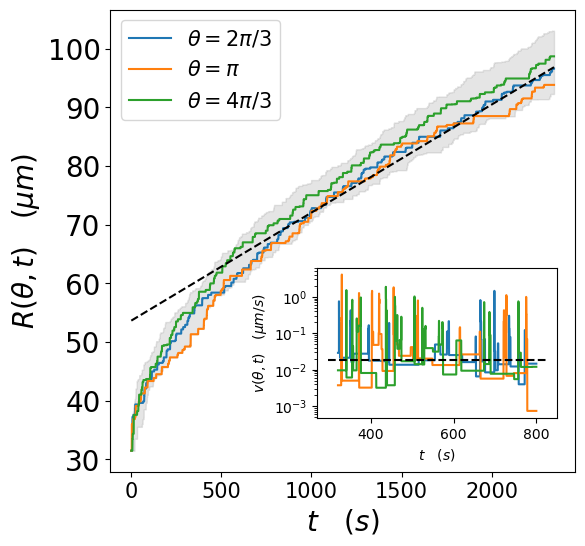}
\caption{
%\textit{Growth  of the domain wall}. 
(a): The evolution of a portion (in pink) of the wall during $8$ sec. The sequence of the frame events is organized in two distinct clusters (in blue and red). The color gradient represents the progression of time, much faster than the steady velocity $\bar{v}$. (b): Time evolution of wall position $R(\theta,t)$ along  three directions (the gray shadow indicates the total spreading). After an initial transient of $t \sim  300 s$, the velocity of the wall decays to its steady value  (e.g. $\bar{v} \sim 0.018 \, \mu m / s $ for $H=0.14$). Inset: Local velocity along three directions. For a given direction $\theta$ we find the times $t_1, t_2, \dots t_i \dots$ where $R(\theta,t)$ changes values. The velocity $v(\theta, t)$ for $t \in [t_i, t_{i+1}]$ is obtained as $(R(\theta, t_{i+1})-R(\theta,t_i))/(t_{i+1}-t_i)$. The dashed line corresponds to the average steady velocity $\bar{v}$. Each signal displays intermittency with instantaneous velocity $100$ larger than $\bar{v}$.}
\label{fig:intf_clusters}
\end{figure*}

In this paper, we report the full dynamical evolution of a domain wall using the magneto-optical Kerr effect (MOKE) on an ultrathin magnetic thin film. As it is clear from the movie in \cite{Note1}, the dynamics is intermittent and correlated. Our analysis demonstrates that the correlations are on scales much larger than $L_{{\rm opt}}$ and that the destabilization and reorganization are governed by the depinning critical point, displaying scale-free statistics with exponents in agreement with the quenched Kardar-Parisi-Zhang (qKPZ) universality class.

\paragraph{Experimental setting. ---}\label{sec:exp}
Field-driven domain wall dynamics is investigated in a  Ta(5)/CoFeB(1)/MgO(2)/Ta(3) (thickness in nm) thin film with perpendicular magnetic anisotropy (PMA)  \footnote{Supplentary Material with additional information about: (S1) Sample preparation and experimental details (S2) Resolving the wall dynamics (S3) Clustering algorithm (S4) Longitudinal length of the cluster (S5) Movie of the experiment at $H=0.13$ mT}. This material is typically very soft, exhibiting a depinning field  of the order of $10$ mT.  The low density of pinning defects with respect to other PMA systems, such as Co/Pt and Co/Ni multilayers, makes it a good candidate to study domain wall dynamics~\cite{BUR-13b}. The competition between domain wall elasticity and the local disorder results in a thermally activated creep motion for driving fields up to the depinning field~\cite{HER-15a}. 
A magnetic bubble domain is initially nucleated with a $\sim$ 30 $\mu$m radius in the pre-saturated film through a short field pulse. The subsequent slow expansion occurs under a small continuous perpendicular applied field. Here we use $H= 0.13, 0.14, 0.15, 0.16$ mT, corresponding to $< 2 \, \% \textrm{ of } H_{\rm dep}$. This ultra-slow creep dynamics is captured through MOKE microscopy. MOKE images with a spatial resolution of $400$ nm are acquired every $200$ ms until the bubble radius has increased to about 100 $\mu$m. Even at the lowest applied field, the bubble domain conserves its circular shape and boundary smoothness upon expansion, indicating weak random pinning. The  limitations in the spatial resolution and in the acquisition rate do not allow us to detect the fast dynamics of the domain wall at the nanoscale, but we can resolve the motion of the wall by estimating the time at which each pixel changes its gray level (see section 1 of \cite{Note1} for a detailed description of the procedure). Remarkably, the set of switched pixels between two consecutive images is always  connected in space, and we define it as a single \textit{frame event}. 
\begin{figure*}[ht]
\centering
 \includegraphics[width=0.75\linewidth]{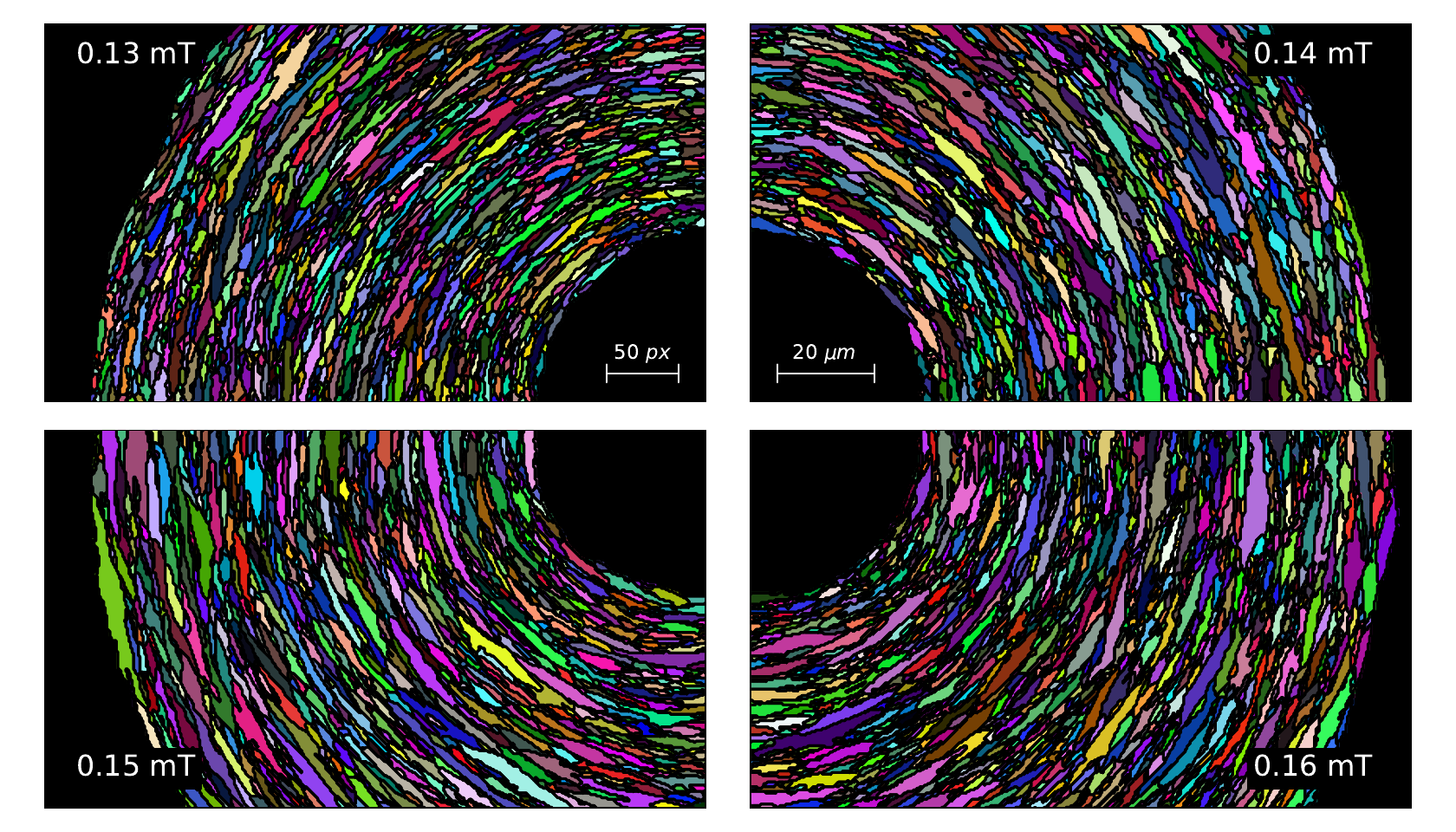}
\caption{Sequences of clusters at different applied fields, which start from the initial bubble (the central black sector at the inner corner of the images) and grow until the radius is about $100\, \mu$m. See also the movie in \cite{Note1}.}
\label{fig:bubbles}
\end{figure*}

\paragraph{Analysis of experimental spatiotemporal patterns. ---}\label{sec:analysis} The dynamics of the domain observed frame by frame displays two important features:
\begin{itemize}
\item  The bubble always expands and never comes back. As shown in Fig.~\ref{fig:intf_clusters} (b), the position of the interface $R(\theta,t)$ along any direction is a non-decreasing function of time. Moreover, after an initial transient, the velocity of the wall decays to its steady value $\bar{v}$. However, the local velocity (inset) displays strong intermittency in time.  
\item The motion presents spatial correlations well above the pixel size. Indeed Fig.~\ref{fig:intf_clusters} (a) shows that each event frame corresponds to a compact spatial region, and events of subsequent frames tend to cluster in space. See the movie in \cite{Note1} to visualize the full dynamics of the bubble.
\end{itemize}

\begin{table}[b]
%\begin{adjustbox}{width=0.4\textwidth}
\begin{tabular}{|c||c||c||c||c|}\hline
    & $\zeta$ & $\tau$ & $1 + 2\zeta$ & $\kappa$\\\hline
    equilibrium &2/3& 4/5 & 7/3 & 2/3\\\hline
    qEW & 1.25 & 1.11 & 3.50 & 1.25\\\hline
    qKPZ & 0.63 & 1.26 & 2.27 & 1.42\\\hline
\end{tabular}
%\end{adjustbox}
%\caption{Critical exponents for depinning  (qEW and qKPZ universality classes)  and equilibrium. Theoretical values of three critical exponents measured in the bubble expansion: The  roughness exponent $\zeta$ ($S \sim \ell^{1+\zeta}$, and $S(q) \sim q^{1+2\zeta}$), the exponent $\tau$ of cluster size distribution $P(S) \sim S^{-\tau}$, and the exponent $\kappa$ for longitudinal length distribution $P(\ell) \sim \ell^{-\kappa} $.}
\caption{Theoretical values of critical exponents for depinning  (qEW and qKPZ universality classes) and equilibrium: the  roughness exponent $\zeta$ ($S \sim \ell^{1+\zeta}$, and $S(q) \sim q^{1+2\zeta}$), the exponent $\tau$ of cluster size distribution $P(S) \sim S^{-\tau}$, and the exponent $\kappa$ for longitudinal length distribution $P(\ell) \sim \ell^{-\kappa} $.}
\label{tab:dep_exponent}
\end{table}

\begin{figure*}[ht]
\textbf{(a)}
\includegraphics[width=0.29\linewidth]{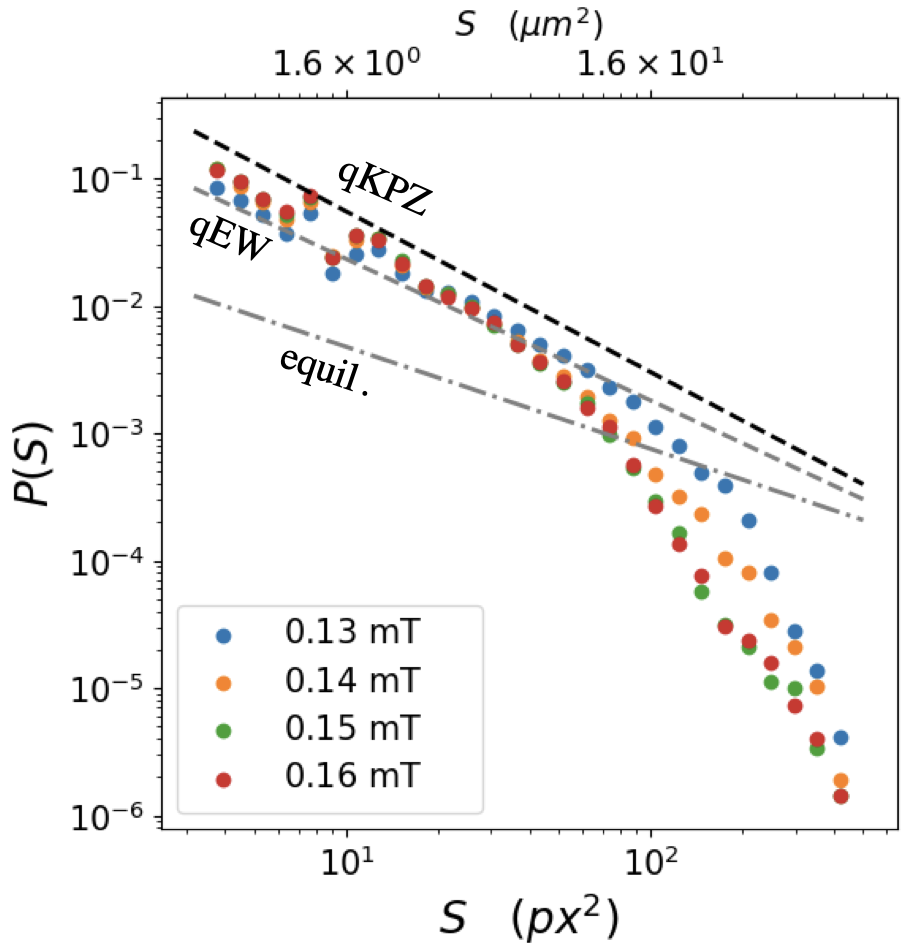} 
\textbf{(b)} %
\includegraphics[width=0.29\linewidth]{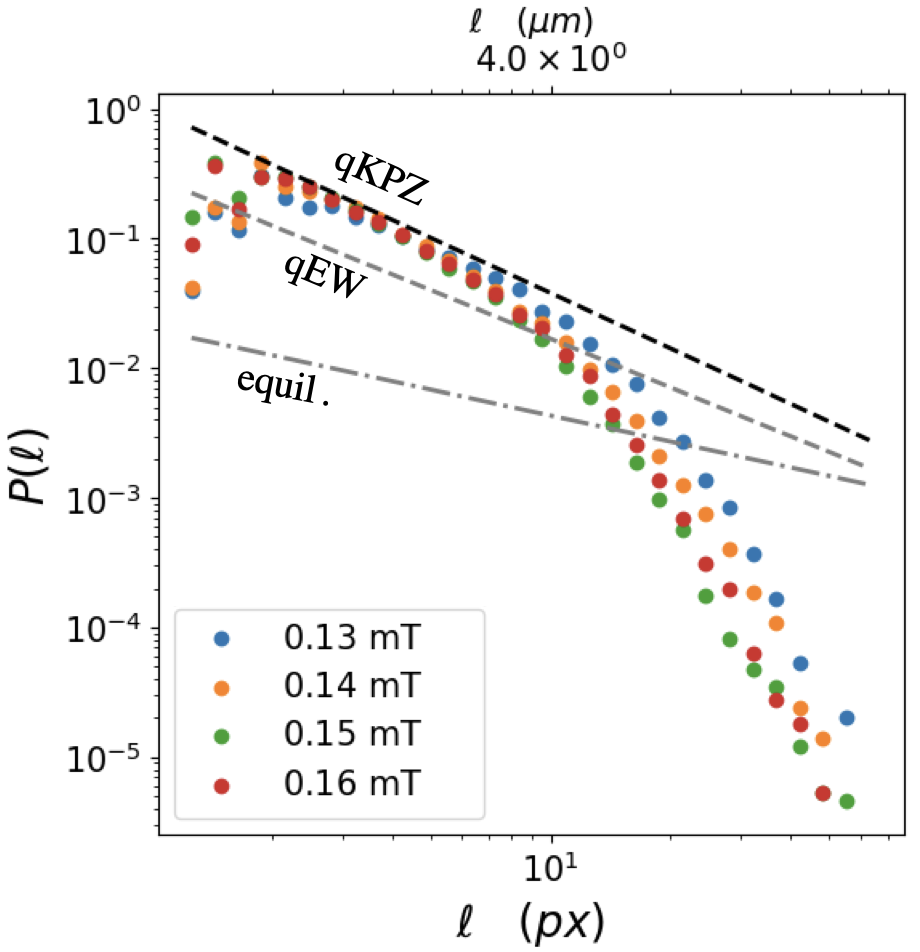} 
\textbf{(c)} %
\includegraphics[width=0.29\linewidth]{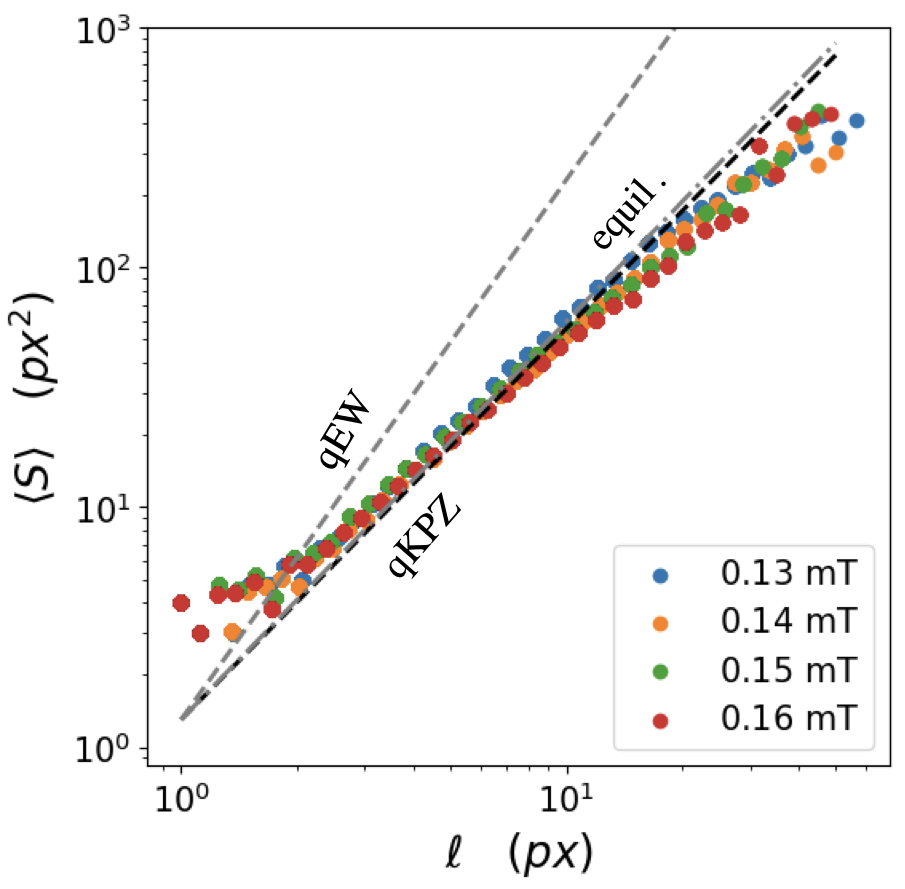}
\caption{
%\textit{Cluster analysis.} 
(a) Cluster size $S$  and (b) longitudinal length $\ell$ distributions for different magnetic fields. (c) Cluster size versus their longitudinal length. The clusters have been obtained for $\Delta t=8$ frames and $\Delta s=2$ pixels. The first two panels   are compatible with qEW and qKPZ universality classes but not with the equilibrium exponents.
The value of the roughness exponents from (c) is computed using the power law scaling $S \sim \ell^{1+\zeta}$. The  measured value is compatible with both $\zeta_{\rm qKPZ}=0.63$ and $\zeta_{\rm equilibrium}=2/3$, but exclude the qEW universality class $\zeta_{\rm qEW}=1.25$. 
Combining these findings leaves the qKPZ universality class as the sole possible candidate for describing the creep motion in our experiment.
}
\label{fig:cluster_stats}
\end{figure*}

\begin{figure}[ht]
\centering
\includegraphics[height=7cm]{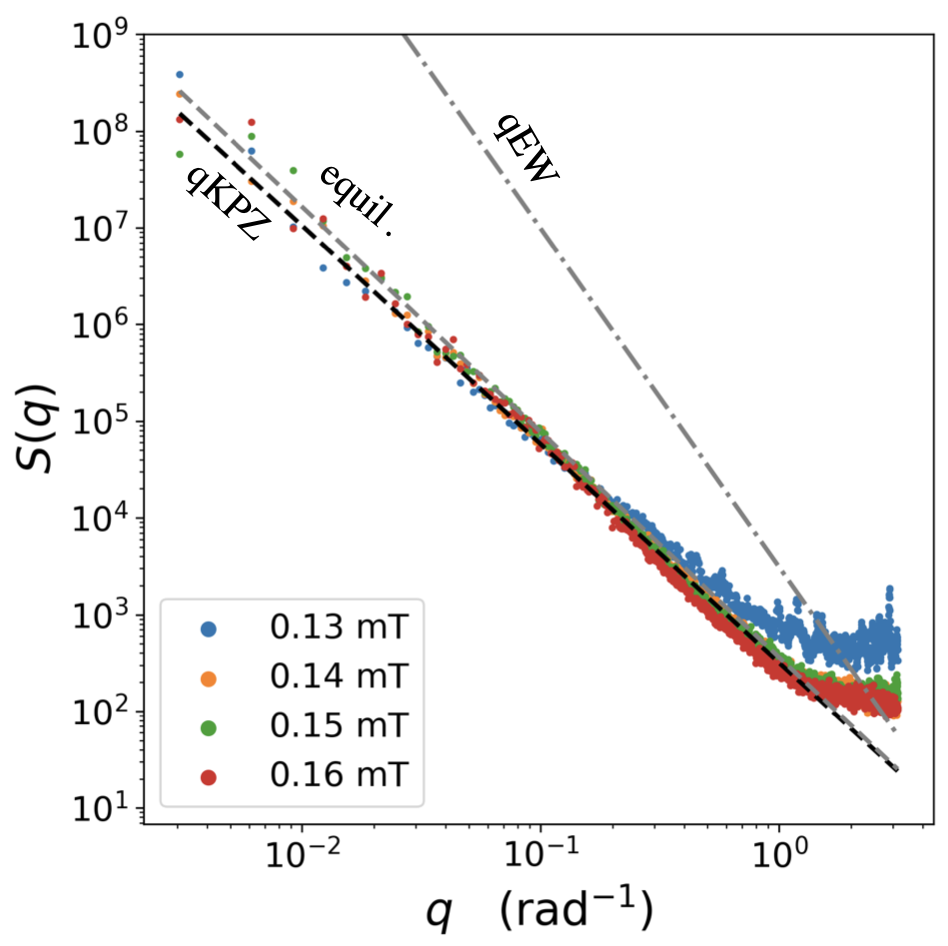}
\caption{
%\textit{Roughness exponent from structure factor.} 
Structure factor $S(q)$ computed  by averaging over time the Fourier transform $\rho(q,t)$ of $\rho(\theta,t) = R(\theta,t) - \overline{R}(t)$, i.e., $S(q) = \langle \rho(q,t) \rho(-q,t) \rangle_t$. 
%The quantity $\rho(\theta,t)$ is obtained from $R(\theta,t)$ by subtracting the average radius $\overline{R}(t)$. 
Using the power law scaling  an $S(q) \sim q^{-(1+2\zeta)}$ at small $q$, we compute a value of the roughness exponent in perfect agreement with the one obtained in Fig.~\ref{fig:cluster_stats} (c). }
\label{fig:roughness}
\end{figure}

These two features support the second scenario for which the initial reorganization of a region of size $L_{\rm opt}$ is followed by a cascade of frame events on much larger scales. Indeed the simple thermal activation is characterized by incoherent back-and-forth motion representing the attempts to overcome the energy barrier. Here, instead, we observe a coherent forward motion on time scales much faster than the steady velocity. This conclusion is also coherent with the estimation of $L_{\rm opt}$ given in \cite{KIM-09, CHA-00}:
\begin{equation}
    L_{\rm opt} \sim L_C (H_{\rm dep}/H)^{3/4}
\end{equation}
with $L_C$ the microscopic Larkin length at which the wall fluctuations become of the order of its thickness. In the materials used in this work, the Larkin length is approximately $L_C \sim 100 \textrm{ nm}$. Hence, $L_{\rm opt}$ is $\sim  380-400\textrm{ nm}$. This scale is just below the single pixel size of $400\textrm{nm}$ and is too small to be experimentally accessible. To quantify the spatial correlations observed beyond $L_{\rm opt}$, we construct clusters of frame events close in space and time via a  simple algorithm that depends on two parameters $\Delta t$ and $\Delta s$. In practice, we start from an initial frame event (the epicenter of the cluster) and include all frame events within a time window $\Delta t$ and a distance $\Delta s$. Section 3 of \cite{Note1} shows that our analysis is robust upon variations of $\Delta t$ and $\Delta s$. Fig.~\ref{fig:bubbles} shows the clusters obtained using this procedure. Each cluster can be characterized by two quantities, namely the size $S$ (the colored areas in Fig.~\ref{fig:bubbles}) and the longitudinal length $\ell$ (see section 4 of \cite{Note1}). Both quantities display scale-free statistics (Fig.~\ref{fig:cluster_stats} (a) and (b)), with exponents which are incompatible with the equilibrium exponents used to characterize the barrier of the energy landscape up to the scale $L_{\rm opt}$. 
It is thus tempting to interpret these clusters as avalanches at the depinning transition as suggested by the numerical simulations on directed interfaces in \cite{FER-17}. In those simulations, however, avalanches are very fat in the growth direction (i.e., the direction of propagation of the interface) consistently with the quenched Edwards Wilkinson (qEW) depinning. Here clusters are instead elongated objects as shown in Figs.~\ref{fig:bubbles} and ~\ref{fig:cluster_stats} (c) where $S \sim \ell^{1+\zeta}$  results in a roughness exponent $\zeta \sim 0.63$. This exponent excludes the possibility of qEW depinning but is consistent with the qKPZ depinning. We corroborate this conclusion with an independent study of the roughness of the whole interface. Following the method proposed in~\cite{TAK-10}, we compute the structure factor $S(q)$ that, as discussed  in Ref.~\cite{FER-17}, displays a $q^{-(1+2\zeta)}$ dependence at small values of the wave number $q$. Fig.~\ref{fig:roughness} shows that the interface's roughness exponent $\zeta$ is consistent with the one characterizing the elongated shape of the cluster. Our results thus prove that the spatial correlations observed beyond the scale $L_{{\rm opt}}$ are in the qKPZ depinning. The qKPZ universality class reveals the presence of anisotropic disorder in our experiment. This feature was not included in previous numerical simulations.

\paragraph{Conclusions. ---}\label{sec:conclusion}
The celebrated creep formula (\ref{eq:creep_formula}) rests on the hypothesis that the key feature determining a wall motion is optimized excitations of size $L_{\rm{opt}}$.
Our work focuses on intermittently occurring rapid movements along a magnetic wall and unveils their spatial organization, extending on scales much more extensive than $L_{\rm{opt}}$. Their size and shape display the same statistics of the avalanches recorded at the depinning magnetic field but with a much slower evolution. In contrast with previous theoretical and experimental studies \cite{FER-21, GRA-18}, our experiment shows that the exponents are compatible with the qKPZ instead of the qEW universality class. The emergence of KPZ dynamics at depinning must be sustained by anisotropy in the material, and its origin calls for further understanding.
%%%Our results support a scenario that seems to be general in several disordered systems, such as amorphous solid or glass-forming liquids \cite{TBOPW-PP-23}. Also, in these systems, the slow dynamics is dominated by localized excitations that trigger a cascade of events displaying scale-free statistics. These thermally-facilitated avalanches are the fingerprint of a complex energy landscape that cannot be reduced to a sequence of uncorrelated elementary excitations.

The scenario emerging from our results should be tested in other examples of elastic disordered systems such as ferroelectric domain walls \cite{TYB-02,CAS-21,TUC-21} or crack propagation \cite{PON-06}. Interestingly, a similar scenario was recently reported for a different class of disordered systems, such as amorphous solids or glass-forming liquids. Simulations on elastoplastic models have shown how localized excitations can trigger cascades of faster events \cite{ OZA-23, TBOPW-PP-23}. Hence, thermally-facilitated avalanches can be pretty generic in disordered systems. They reveal the complex nature of disordered energy landscapes that cannot be described simply by a sequence of uncorrelated elementary excitations.

The results reported here can also have significant consequences in the field of spintronics. The creep dynamics of a bubble domain is, in fact, at the base of one of the most used methods to determine the interfacial Dzyaloshinskii-Moriya interaction (DMI). This is a chiral interaction responsible for the occurrence of topological spin structures, such as chiral domain walls and skyrmions, considered the most promising information carriers in future spintronics technologies \cite{KUE-23}. The determination of the DMI constant is based on the asymmetric expansion of the bubble under an in-plane magnetic field, with the domain wall velocity measured by dividing the displacement between two MOKE snapshots over their time interval. Fig.~\ref{fig:intf_clusters} (b) actually suggests that the velocity is constant only at large times/displacements, and thus that this procedure could be misleading. In addition, theoretical expressions to evaluate the DMI field from the velocity curve are primarily phenomenological, and a more accurate description of the domain wall dynamics, such as the qKPZ reported here, could highly improve the fits of the data. We hope these considerations shed some light on a more accurate determination of DMI value and solve the contradictions with other popular methods, such as the Brillouin light scattering. 

\paragraph{Acknowledgements. ---} V.M.S. acknowledges 80Prime
CNRS support for the project CorrQuake.
M.B. is supported by research grant BAIE$\_$BIRD2021$\_$01 of the University of Padova.

%%%% SUPPL. %%%%%

%


\begin{thebibliography}{31}%
	\makeatletter
	\providecommand \@ifxundefined [1]{%
		\@ifx{#1\undefined}
	}%
	\providecommand \@ifnum [1]{%
		\ifnum #1\expandafter \@firstoftwo
		\else \expandafter \@secondoftwo
		\fi
	}%
	\providecommand \@ifx [1]{%
		\ifx #1\expandafter \@firstoftwo
		\else \expandafter \@secondoftwo
		\fi
	}%
	\providecommand \natexlab [1]{#1}%
	\providecommand \enquote  [1]{``#1''}%
	\providecommand \bibnamefont  [1]{#1}%
	\providecommand \bibfnamefont [1]{#1}%
	\providecommand \citenamefont [1]{#1}%
	\providecommand \href@noop [0]{\@secondoftwo}%
	\providecommand \href [0]{\begingroup \@sanitize@url \@href}%
	\providecommand \@href[1]{\@@startlink{#1}\@@href}%
	\providecommand \@@href[1]{\endgroup#1\@@endlink}%
	\providecommand \@sanitize@url [0]{\catcode `\\12\catcode `\$12\catcode
		`\&12\catcode `\#12\catcode `\^12\catcode `\_12\catcode `\%12\relax}%
	\providecommand \@@startlink[1]{}%
	\providecommand \@@endlink[0]{}%
	\providecommand \url  [0]{\begingroup\@sanitize@url \@url }%
	\providecommand \@url [1]{\endgroup\@href {#1}{\urlprefix }}%
	\providecommand \urlprefix  [0]{URL }%
	\providecommand \Eprint [0]{\href }%
	\providecommand \doibase [0]{https://doi.org/}%
	\providecommand \selectlanguage [0]{\@gobble}%
	\providecommand \bibinfo  [0]{\@secondoftwo}%
	\providecommand \bibfield  [0]{\@secondoftwo}%
	\providecommand \translation [1]{[#1]}%
	\providecommand \BibitemOpen [0]{}%
	\providecommand \bibitemStop [0]{}%
	\providecommand \bibitemNoStop [0]{.\EOS\space}%
	\providecommand \EOS [0]{\spacefactor3000\relax}%
	\providecommand \BibitemShut  [1]{\csname bibitem#1\endcsname}%
	\let\auto@bib@innerbib\@empty
	%</preamble>
	\bibitem [{\citenamefont {Parkin}\ and\ \citenamefont {Yang}(2015)}]{PAR-15}%
	\BibitemOpen
	\bibfield  {author} {\bibinfo {author} {\bibfnamefont {S.}~\bibnamefont
			{Parkin}}\ and\ \bibinfo {author} {\bibfnamefont {S.-H.}\ \bibnamefont
			{Yang}},\ }\bibfield  {title} {\bibinfo {title} {Memory on the racetrack},\
	}\href {https://doi.org/10.1038/nnano.2015.41} {\bibfield  {journal}
		{\bibinfo  {journal} {Nature Nanotechnology}\ }\textbf {\bibinfo {volume}
			{10}},\ \bibinfo {pages} {195} (\bibinfo {year} {2015})}\BibitemShut
	{NoStop}%
	\bibitem [{\citenamefont {Gu}\ \emph {et~al.}(2022)\citenamefont {Gu},
		\citenamefont {Guan}, \citenamefont {Hazra}, \citenamefont {Deniz},
		\citenamefont {Migliorini}, \citenamefont {Zhang},\ and\ \citenamefont
		{Parkin}}]{GU-22}%
	\BibitemOpen
	\bibfield  {author} {\bibinfo {author} {\bibfnamefont {K.}~\bibnamefont
			{Gu}}, \bibinfo {author} {\bibfnamefont {Y.}~\bibnamefont {Guan}}, \bibinfo
		{author} {\bibfnamefont {B.~K.}\ \bibnamefont {Hazra}}, \bibinfo {author}
		{\bibfnamefont {H.}~\bibnamefont {Deniz}}, \bibinfo {author} {\bibfnamefont
			{A.}~\bibnamefont {Migliorini}}, \bibinfo {author} {\bibfnamefont
			{W.}~\bibnamefont {Zhang}},\ and\ \bibinfo {author} {\bibfnamefont
			{S.~S.~P.}\ \bibnamefont {Parkin}},\ }\bibfield  {title} {\bibinfo {title}
		{Three-dimensional racetrack memory devices designed from freestanding
			magnetic heterostructures},\ }\href
	{https://doi.org/10.1038/s41565-022-01213-1} {\bibfield  {journal} {\bibinfo
			{journal} {Nature Nanotechnology}\ }\textbf {\bibinfo {volume} {17}},\
		\bibinfo {pages} {1065} (\bibinfo {year} {2022})}\BibitemShut {NoStop}%
	\bibitem [{\citenamefont {Ioffe}\ and\ \citenamefont {Vinokur}(1987)}]{IV-87}%
	\BibitemOpen
	\bibfield  {author} {\bibinfo {author} {\bibfnamefont {L.~B.}\ \bibnamefont
			{Ioffe}}\ and\ \bibinfo {author} {\bibfnamefont {V.~M.}\ \bibnamefont
			{Vinokur}},\ }\bibfield  {title} {\bibinfo {title} {Dynamics of interfaces
			and dislocations in disordered media},\ }\href
	{https://doi.org/10.1088/0022-3719/20/36/016} {\bibfield  {journal} {\bibinfo
			{journal} {Journal of Physics C: Solid State Physics}\ }\textbf {\bibinfo
			{volume} {20}},\ \bibinfo {pages} {6149} (\bibinfo {year}
		{1987})}\BibitemShut {NoStop}%
	\bibitem [{\citenamefont {Lemerle}\ \emph {et~al.}(1998)\citenamefont
		{Lemerle}, \citenamefont {Ferr\'e}, \citenamefont {Chappert}, \citenamefont
		{Mathet}, \citenamefont {Giamarchi},\ and\ \citenamefont
		{Le~Doussal}}]{LEM-98}%
	\BibitemOpen
	\bibfield  {author} {\bibinfo {author} {\bibfnamefont {S.}~\bibnamefont
			{Lemerle}}, \bibinfo {author} {\bibfnamefont {J.}~\bibnamefont {Ferr\'e}},
		\bibinfo {author} {\bibfnamefont {C.}~\bibnamefont {Chappert}}, \bibinfo
		{author} {\bibfnamefont {V.}~\bibnamefont {Mathet}}, \bibinfo {author}
		{\bibfnamefont {T.}~\bibnamefont {Giamarchi}},\ and\ \bibinfo {author}
		{\bibfnamefont {P.}~\bibnamefont {Le~Doussal}},\ }\bibfield  {title}
	{\bibinfo {title} {Domain wall creep in an {I}sing ultrathin magnetic film},\
	}\href {https://doi.org/10.1103/PhysRevLett.80.849} {\bibfield  {journal}
		{\bibinfo  {journal} {Phys. Rev. Lett.}\ }\textbf {\bibinfo {volume} {80}},\
		\bibinfo {pages} {849} (\bibinfo {year} {1998})}\BibitemShut {NoStop}%
	\bibitem [{KIM(2009)}]{KIM-09}%
	\BibitemOpen
	\bibfield  {title} {\bibinfo {title} {Interdimensional universality of
			dynamic interfaces},\ }\href {https://doi.org/10.1038/nature07874} {\bibfield
		{journal} {\bibinfo  {journal} {Nature}\ }\textbf {\bibinfo {volume}
			{458}},\ \bibinfo {pages} {740} (\bibinfo {year} {2009})}\BibitemShut
	{NoStop}%
	\bibitem [{\citenamefont {Jeudy}\ \emph {et~al.}(2016)\citenamefont {Jeudy},
		\citenamefont {Mougin}, \citenamefont {Bustingorry}, \citenamefont
		{Savero~Torres}, \citenamefont {Gorchon}, \citenamefont {Kolton},
		\citenamefont {Lemaître},\ and\ \citenamefont {Jamet}}]{JEU-16}%
	\BibitemOpen
	\bibfield  {author} {\bibinfo {author} {\bibfnamefont {V.}~\bibnamefont
			{Jeudy}}, \bibinfo {author} {\bibfnamefont {A.}~\bibnamefont {Mougin}},
		\bibinfo {author} {\bibfnamefont {S.}~\bibnamefont {Bustingorry}}, \bibinfo
		{author} {\bibfnamefont {W.}~\bibnamefont {Savero~Torres}}, \bibinfo {author}
		{\bibfnamefont {J.}~\bibnamefont {Gorchon}}, \bibinfo {author} {\bibfnamefont
			{A.~B.}\ \bibnamefont {Kolton}}, \bibinfo {author} {\bibfnamefont
			{A.}~\bibnamefont {Lemaître}},\ and\ \bibinfo {author} {\bibfnamefont
			{J.-P.}\ \bibnamefont {Jamet}},\ }\bibfield  {title} {\bibinfo {title}
		{Universal pinning energy barrier for driven domain walls in thin
			ferromagnetic films},\ }\href
	{https://doi.org/10.1103/physrevlett.117.057201} {\bibfield  {journal}
		{\bibinfo  {journal} {Phys. Rev. Lett.}\ }\textbf {\bibinfo {volume} {117}},\
		\bibinfo {pages} {057201} (\bibinfo {year} {2016})}\BibitemShut {NoStop}%
	\bibitem [{\citenamefont {Dong}\ \emph {et~al.}(1993)\citenamefont {Dong},
		\citenamefont {Marchetti}, \citenamefont {Middleton},\ and\ \citenamefont
		{Vinokur}}]{DMMAV}%
	\BibitemOpen
	\bibfield  {author} {\bibinfo {author} {\bibfnamefont {M.}~\bibnamefont
			{Dong}}, \bibinfo {author} {\bibfnamefont {M.}~\bibnamefont {Marchetti}},
		\bibinfo {author} {\bibfnamefont {A.~A.}\ \bibnamefont {Middleton}},\ and\
		\bibinfo {author} {\bibfnamefont {V.}~\bibnamefont {Vinokur}},\ }\bibfield
	{title} {\bibinfo {title} {Elastic string in a random potential},\ }\href
	{https://doi.org/10.1103/PhysRevLett.70.662} {\bibfield  {journal} {\bibinfo
			{journal} {Phys. Rev. Lett.}\ }\textbf {\bibinfo {volume} {70}},\ \bibinfo
		{pages} {662} (\bibinfo {year} {1993})}\BibitemShut {NoStop}%
	\bibitem [{\citenamefont {Agoritsas}\ \emph {et~al.}(2012)\citenamefont
		{Agoritsas}, \citenamefont {Lecomte},\ and\ \citenamefont
		{Giamarchi}}]{AGO-12}%
	\BibitemOpen
	\bibfield  {author} {\bibinfo {author} {\bibfnamefont {E.}~\bibnamefont
			{Agoritsas}}, \bibinfo {author} {\bibfnamefont {V.}~\bibnamefont {Lecomte}},\
		and\ \bibinfo {author} {\bibfnamefont {T.}~\bibnamefont {Giamarchi}},\
	}\bibfield  {title} {\bibinfo {title} {Disordered elastic systems and
			one-dimensional interfaces},\ }\href
	{https://doi.org/http://dx.doi.org/10.1016/j.physb.2012.01.017} {\bibfield
		{journal} {\bibinfo  {journal} {Physica B: Condensed Matter}\ }\textbf
		{\bibinfo {volume} {407}},\ \bibinfo {pages} {1725 } (\bibinfo {year}
		{2012})},\ \bibinfo {note} {proceedings of the International Workshop on
		Electronic Crystals (ECRYS-2011)}\BibitemShut {NoStop}%
	\bibitem [{\citenamefont {Ferrero}\ \emph {et~al.}(2021)\citenamefont
		{Ferrero}, \citenamefont {Foini}, \citenamefont {Giamarchi}, \citenamefont
		{Kolton},\ and\ \citenamefont {Rosso}}]{FER-21}%
	\BibitemOpen
	\bibfield  {author} {\bibinfo {author} {\bibfnamefont {E.~E.}\ \bibnamefont
			{Ferrero}}, \bibinfo {author} {\bibfnamefont {L.}~\bibnamefont {Foini}},
		\bibinfo {author} {\bibfnamefont {T.}~\bibnamefont {Giamarchi}}, \bibinfo
		{author} {\bibfnamefont {A.~B.}\ \bibnamefont {Kolton}},\ and\ \bibinfo
		{author} {\bibfnamefont {A.}~\bibnamefont {Rosso}},\ }\bibfield  {title}
	{\bibinfo {title} {Creep motion of elastic interfaces driven in a disordered
			landscape},\ }\href@noop {} {\bibfield  {journal} {\bibinfo  {journal}
			{Annual Review of Condensed Matter Physics}\ }\textbf {\bibinfo {volume}
			{12}},\ \bibinfo {pages} {111} (\bibinfo {year} {2021})}\BibitemShut
	{NoStop}%
	\bibitem [{\citenamefont {Chauve}\ \emph {et~al.}(2000)\citenamefont {Chauve},
		\citenamefont {Giamarchi},\ and\ \citenamefont {Le~Doussal}}]{CHA-00}%
	\BibitemOpen
	\bibfield  {author} {\bibinfo {author} {\bibfnamefont {P.}~\bibnamefont
			{Chauve}}, \bibinfo {author} {\bibfnamefont {T.}~\bibnamefont {Giamarchi}},\
		and\ \bibinfo {author} {\bibfnamefont {P.}~\bibnamefont {Le~Doussal}},\
	}\bibfield  {title} {\bibinfo {title} {Creep and depinning in disordered
			media},\ }\href {https://doi.org/10.1103/PhysRevB.62.6241} {\bibfield
		{journal} {\bibinfo  {journal} {Phys. Rev. B}\ }\textbf {\bibinfo {volume}
			{62}},\ \bibinfo {pages} {6241} (\bibinfo {year} {2000})}\BibitemShut
	{NoStop}%
	\bibitem [{\citenamefont {Kolton}\ \emph {et~al.}(2009)\citenamefont {Kolton},
		\citenamefont {Rosso}, \citenamefont {Giamarchi},\ and\ \citenamefont
		{Krauth}}]{KOL-09}%
	\BibitemOpen
	\bibfield  {author} {\bibinfo {author} {\bibfnamefont {A.~B.}\ \bibnamefont
			{Kolton}}, \bibinfo {author} {\bibfnamefont {A.}~\bibnamefont {Rosso}},
		\bibinfo {author} {\bibfnamefont {T.}~\bibnamefont {Giamarchi}},\ and\
		\bibinfo {author} {\bibfnamefont {W.}~\bibnamefont {Krauth}},\ }\bibfield
	{title} {\bibinfo {title} {Creep dynamics of elastic manifolds via exact
			transition pathways},\ }\href {https://doi.org/10.1103/PhysRevB.79.184207}
	{\bibfield  {journal} {\bibinfo  {journal} {Phys. Rev. B}\ }\textbf {\bibinfo
			{volume} {79}},\ \bibinfo {pages} {184207} (\bibinfo {year}
		{2009})}\BibitemShut {NoStop}%
	\bibitem [{\citenamefont {Ferrero}\ \emph {et~al.}(2017)\citenamefont
		{Ferrero}, \citenamefont {Foini}, \citenamefont {Giamarchi}, \citenamefont
		{Kolton},\ and\ \citenamefont {Rosso}}]{FER-17}%
	\BibitemOpen
	\bibfield  {author} {\bibinfo {author} {\bibfnamefont {E.~E.}\ \bibnamefont
			{Ferrero}}, \bibinfo {author} {\bibfnamefont {L.}~\bibnamefont {Foini}},
		\bibinfo {author} {\bibfnamefont {T.}~\bibnamefont {Giamarchi}}, \bibinfo
		{author} {\bibfnamefont {A.~B.}\ \bibnamefont {Kolton}},\ and\ \bibinfo
		{author} {\bibfnamefont {A.}~\bibnamefont {Rosso}},\ }\bibfield  {title}
	{\bibinfo {title} {Spatiotemporal patterns in ultraslow domain wall creep
			dynamics},\ }\href {https://doi.org/10.1103/PhysRevLett.118.147208}
	{\bibfield  {journal} {\bibinfo  {journal} {Phys. Rev. Lett.}\ }\textbf
		{\bibinfo {volume} {118}},\ \bibinfo {pages} {147208} (\bibinfo {year}
		{2017})}\BibitemShut {NoStop}%
	\bibitem [{\citenamefont {Purrello}\ \emph {et~al.}(2017)\citenamefont
		{Purrello}, \citenamefont {Iguain}, \citenamefont {Kolton},\ and\
		\citenamefont {Jagla}}]{PUR-17}%
	\BibitemOpen
	\bibfield  {author} {\bibinfo {author} {\bibfnamefont {V.~H.}\ \bibnamefont
			{Purrello}}, \bibinfo {author} {\bibfnamefont {J.~L.}\ \bibnamefont
			{Iguain}}, \bibinfo {author} {\bibfnamefont {A.~B.}\ \bibnamefont {Kolton}},\
		and\ \bibinfo {author} {\bibfnamefont {E.~A.}\ \bibnamefont {Jagla}},\
	}\bibfield  {title} {\bibinfo {title} {Creep and thermal rounding close to
			the elastic depinning threshold},\ }\href
	{https://doi.org/10.1103/PhysRevE.96.022112} {\bibfield  {journal} {\bibinfo
			{journal} {Phys. Rev. E}\ }\textbf {\bibinfo {volume} {96}},\ \bibinfo
		{pages} {022112} (\bibinfo {year} {2017})}\BibitemShut {NoStop}%
	\bibitem [{\citenamefont {Baiesi}\ and\ \citenamefont
		{Paczuski}(2004)}]{BAI-04}%
	\BibitemOpen
	\bibfield  {author} {\bibinfo {author} {\bibfnamefont {M.}~\bibnamefont
			{Baiesi}}\ and\ \bibinfo {author} {\bibfnamefont {M.}~\bibnamefont
			{Paczuski}},\ }\bibfield  {title} {\bibinfo {title} {Scale-free networks of
			earthquakes and aftershocks},\ }\bibfield  {journal} {\bibinfo  {journal}
		{Phys. Rev. E}\ }\textbf {\bibinfo {volume} {69}},\ \href
	{https://doi.org/10.1103/physreve.69.066106} {10.1103/physreve.69.066106}
	(\bibinfo {year} {2004})\BibitemShut {NoStop}%
	\bibitem [{\citenamefont {Jagla}\ and\ \citenamefont {Kolton}(2010)}]{JAG-10}%
	\BibitemOpen
	\bibfield  {author} {\bibinfo {author} {\bibfnamefont {E.~A.}\ \bibnamefont
			{Jagla}}\ and\ \bibinfo {author} {\bibfnamefont {A.~B.}\ \bibnamefont
			{Kolton}},\ }\bibfield  {title} {\bibinfo {title} {A mechanism for spatial
			and temporal earthquake clustering},\ }\bibfield  {journal} {\bibinfo
		{journal} {Journal of Geophysical Research}\ }\textbf {\bibinfo {volume}
		{115}},\ \href {https://doi.org/10.1029/2009jb006974} {10.1029/2009jb006974}
	(\bibinfo {year} {2010})\BibitemShut {NoStop}%
	\bibitem [{\citenamefont {Jagla}\ \emph {et~al.}(2014)\citenamefont {Jagla},
		\citenamefont {Landes},\ and\ \citenamefont {Rosso}}]{JAG-14}%
	\BibitemOpen
	\bibfield  {author} {\bibinfo {author} {\bibfnamefont {E.~A.}\ \bibnamefont
			{Jagla}}, \bibinfo {author} {\bibfnamefont {F.~m. c.~P.}\ \bibnamefont
			{Landes}},\ and\ \bibinfo {author} {\bibfnamefont {A.}~\bibnamefont
			{Rosso}},\ }\bibfield  {title} {\bibinfo {title} {Viscoelastic effects in
			avalanche dynamics: A key to earthquake statistics},\ }\href
	{https://doi.org/10.1103/PhysRevLett.112.174301} {\bibfield  {journal}
		{\bibinfo  {journal} {Phys. Rev. Lett.}\ }\textbf {\bibinfo {volume} {112}},\
		\bibinfo {pages} {174301} (\bibinfo {year} {2014})}\BibitemShut {NoStop}%
	\bibitem [{\citenamefont {Scholz}(2019)}]{SCH-19}%
	\BibitemOpen
	\bibfield  {author} {\bibinfo {author} {\bibfnamefont {C.~H.}\ \bibnamefont
			{Scholz}},\ }\href@noop {} {\emph {\bibinfo {title} {The mechanics of
				earthquakes and faulting}}}\ (\bibinfo  {publisher} {Cambridge university
		press},\ \bibinfo {year} {2019})\BibitemShut {NoStop}%
	\bibitem [{\citenamefont {Fisher}(1998)}]{FIS-98}%
	\BibitemOpen
	\bibfield  {author} {\bibinfo {author} {\bibfnamefont {D.~S.}\ \bibnamefont
			{Fisher}},\ }\bibfield  {title} {\bibinfo {title} {Collective transport in
			random media: from superconductors to earthquakes},\ }\href
	{https://doi.org/10.1016/s0370-1573(98)00008-8} {\bibfield  {journal}
		{\bibinfo  {journal} {Physics Reports}\ }\textbf {\bibinfo {volume} {301}},\
		\bibinfo {pages} {113} (\bibinfo {year} {1998})}\BibitemShut {NoStop}%
	\bibitem [{\citenamefont {Kardar}(1998)}]{KAR-98}%
	\BibitemOpen
	\bibfield  {author} {\bibinfo {author} {\bibfnamefont {M.}~\bibnamefont
			{Kardar}},\ }\bibfield  {title} {\bibinfo {title} {Nonequilibrium dynamics of
			interfaces and lines},\ }\href
	{https://doi.org/10.1016/s0370-1573(98)00007-6} {\bibfield  {journal}
		{\bibinfo  {journal} {Physics Reports}\ }\textbf {\bibinfo {volume} {301}},\
		\bibinfo {pages} {85} (\bibinfo {year} {1998})}\BibitemShut {NoStop}%
	\bibitem [{\citenamefont {Ozawa}\ and\ \citenamefont {Biroli}(2023)}]{OZA-23}%
	\BibitemOpen
	\bibfield  {author} {\bibinfo {author} {\bibfnamefont {M.}~\bibnamefont
			{Ozawa}}\ and\ \bibinfo {author} {\bibfnamefont {G.}~\bibnamefont {Biroli}},\
	}\bibfield  {title} {\bibinfo {title} {Elasticity, facilitation, and dynamic
			heterogeneity in glass-forming liquids},\ }\href
	{https://doi.org/10.1103/PhysRevLett.130.138201} {\bibfield  {journal}
		{\bibinfo  {journal} {Phys. Rev. Lett.}\ }\textbf {\bibinfo {volume} {130}},\
		\bibinfo {pages} {138201} (\bibinfo {year} {2023})}\BibitemShut {NoStop}%
	\bibitem [{\citenamefont {Tahaei}\ \emph {et~al.}(2023)\citenamefont {Tahaei},
		\citenamefont {Biroli}, \citenamefont {Ozawa}, \citenamefont {Popovi{\'c}},\
		and\ \citenamefont {Wyart}}]{TBOPW-PP-23}%
	\BibitemOpen
	\bibfield  {author} {\bibinfo {author} {\bibfnamefont {A.}~\bibnamefont
			{Tahaei}}, \bibinfo {author} {\bibfnamefont {G.}~\bibnamefont {Biroli}},
		\bibinfo {author} {\bibfnamefont {M.}~\bibnamefont {Ozawa}}, \bibinfo
		{author} {\bibfnamefont {M.}~\bibnamefont {Popovi{\'c}}},\ and\ \bibinfo
		{author} {\bibfnamefont {M.}~\bibnamefont {Wyart}},\ }\bibfield  {title}
	{\bibinfo {title} {Scaling description of dynamical heterogeneity and
			avalanches of relaxation in glass-forming liquids},\ }\bibfield  {journal}
	{\bibinfo  {journal} {arXiv preprint arXiv:2305.00219}\ }\href
	{https://doi.org/10.48550/ARXIV.2305.00219} {10.48550/ARXIV.2305.00219}
	(\bibinfo {year} {2023})\BibitemShut {NoStop}%
	\bibitem [{Note1()}]{Note1}%
	\BibitemOpen
	\bibinfo {note} {Supplentary Material with additional information about: (S1)
		Sample preparation and experimental details (S2) Resolving the wall dynamics
		(S3) Clustering algorithm (S4) Longitudinal length of the cluster (S5) Movie
		of the experiment at $H=0.13$ mT}\BibitemShut {NoStop}%
	\bibitem [{\citenamefont {Burrowes}\ \emph {et~al.}(2013)\citenamefont
		{Burrowes}, \citenamefont {Vernier}, \citenamefont {Adam}, \citenamefont
		{Herrera~Diez}, \citenamefont {Garcia}, \citenamefont {Barisic},
		\citenamefont {Agnus}, \citenamefont {Eimer}, \citenamefont {Kim},
		\citenamefont {Devolder},\ and\ \citenamefont {et~al.}}]{BUR-13b}%
	\BibitemOpen
	\bibfield  {author} {\bibinfo {author} {\bibfnamefont {C.}~\bibnamefont
			{Burrowes}}, \bibinfo {author} {\bibfnamefont {N.}~\bibnamefont {Vernier}},
		\bibinfo {author} {\bibfnamefont {J.-P.}\ \bibnamefont {Adam}}, \bibinfo
		{author} {\bibfnamefont {L.}~\bibnamefont {Herrera~Diez}}, \bibinfo {author}
		{\bibfnamefont {K.}~\bibnamefont {Garcia}}, \bibinfo {author} {\bibfnamefont
			{I.}~\bibnamefont {Barisic}}, \bibinfo {author} {\bibfnamefont
			{G.}~\bibnamefont {Agnus}}, \bibinfo {author} {\bibfnamefont
			{S.}~\bibnamefont {Eimer}}, \bibinfo {author} {\bibfnamefont {J.-V.}\
			\bibnamefont {Kim}}, \bibinfo {author} {\bibfnamefont {T.}~\bibnamefont
			{Devolder}},\ and\ \bibinfo {author} {\bibnamefont {et~al.}},\ }\bibfield
	{title} {\bibinfo {title} {Low depinning fields in ta-cofeb-mgo ultrathin
			films with perpendicular magnetic anisotropy},\ }\href
	{https://doi.org/10.1063/1.4826439} {\bibfield  {journal} {\bibinfo
			{journal} {Appl. Phys. Lett.}\ }\textbf {\bibinfo {volume} {103}},\ \bibinfo
		{pages} {182401} (\bibinfo {year} {2013})}\BibitemShut {NoStop}%
	\bibitem [{\citenamefont {Herrera~Diez}\ \emph {et~al.}(2015)\citenamefont
		{Herrera~Diez}, \citenamefont {García-Sánchez}, \citenamefont {Adam},
		\citenamefont {Devolder}, \citenamefont {Eimer}, \citenamefont {El~Hadri},
		\citenamefont {Lamperti}, \citenamefont {Mantovan}, \citenamefont {Ocker},\
		and\ \citenamefont {Ravelosona}}]{HER-15a}%
	\BibitemOpen
	\bibfield  {author} {\bibinfo {author} {\bibfnamefont {L.}~\bibnamefont
			{Herrera~Diez}}, \bibinfo {author} {\bibfnamefont {F.}~\bibnamefont
			{García-Sánchez}}, \bibinfo {author} {\bibfnamefont {J.-P.}\ \bibnamefont
			{Adam}}, \bibinfo {author} {\bibfnamefont {T.}~\bibnamefont {Devolder}},
		\bibinfo {author} {\bibfnamefont {S.}~\bibnamefont {Eimer}}, \bibinfo
		{author} {\bibfnamefont {M.~S.}\ \bibnamefont {El~Hadri}}, \bibinfo {author}
		{\bibfnamefont {A.}~\bibnamefont {Lamperti}}, \bibinfo {author}
		{\bibfnamefont {R.}~\bibnamefont {Mantovan}}, \bibinfo {author}
		{\bibfnamefont {B.}~\bibnamefont {Ocker}},\ and\ \bibinfo {author}
		{\bibfnamefont {D.}~\bibnamefont {Ravelosona}},\ }\bibfield  {title}
	{\bibinfo {title} {Controlling magnetic domain wall motion in the creep
			regime in {He$^+$}-irradiated {CoFeB/MgO} films with perpendicular
			anisotropy},\ }\href {https://doi.org/10.1063/1.4927204} {\bibfield
		{journal} {\bibinfo  {journal} {Appl. Phys. Lett.}\ }\textbf {\bibinfo
			{volume} {107}},\ \bibinfo {pages} {032401} (\bibinfo {year}
		{2015})}\BibitemShut {NoStop}%
	\bibitem [{\citenamefont {Takeuchi}\ and\ \citenamefont {Sano}(2010)}]{TAK-10}%
	\BibitemOpen
	\bibfield  {author} {\bibinfo {author} {\bibfnamefont {K.~A.}\ \bibnamefont
			{Takeuchi}}\ and\ \bibinfo {author} {\bibfnamefont {M.}~\bibnamefont
			{Sano}},\ }\bibfield  {title} {\bibinfo {title} {Universal fluctuations of
			growing interfaces: Evidence in turbulent liquid crystals},\ }\href
	{https://doi.org/10.1103/physrevlett.104.230601} {\bibfield  {journal}
		{\bibinfo  {journal} {Phys. Rev. Lett.}\ }\textbf {\bibinfo {volume} {104}},\
		\bibinfo {pages} {230601} (\bibinfo {year} {2010})}\BibitemShut {NoStop}%
	\bibitem [{\citenamefont {Grassi}\ \emph {et~al.}(2018)\citenamefont {Grassi},
		\citenamefont {Kolton}, \citenamefont {Jeudy}, \citenamefont {Mougin},
		\citenamefont {Bustingorry},\ and\ \citenamefont {Curiale}}]{GRA-18}%
	\BibitemOpen
	\bibfield  {author} {\bibinfo {author} {\bibfnamefont {M.~P.}\ \bibnamefont
			{Grassi}}, \bibinfo {author} {\bibfnamefont {A.~B.}\ \bibnamefont {Kolton}},
		\bibinfo {author} {\bibfnamefont {V.}~\bibnamefont {Jeudy}}, \bibinfo
		{author} {\bibfnamefont {A.}~\bibnamefont {Mougin}}, \bibinfo {author}
		{\bibfnamefont {S.}~\bibnamefont {Bustingorry}},\ and\ \bibinfo {author}
		{\bibfnamefont {J.}~\bibnamefont {Curiale}},\ }\bibfield  {title} {\bibinfo
		{title} {Intermittent collective dynamics of domain walls in the creep
			regime},\ }\href {https://doi.org/10.1103/PhysRevB.98.224201} {\bibfield
		{journal} {\bibinfo  {journal} {Phys. Rev. B}\ }\textbf {\bibinfo {volume}
			{98}},\ \bibinfo {pages} {224201} (\bibinfo {year} {2018})}\BibitemShut
	{NoStop}%
	\bibitem [{\citenamefont {Tybell}\ \emph {et~al.}(2002)\citenamefont {Tybell},
		\citenamefont {Paruch}, \citenamefont {Giamarchi},\ and\ \citenamefont
		{Triscone}}]{TYB-02}%
	\BibitemOpen
	\bibfield  {author} {\bibinfo {author} {\bibfnamefont {T.}~\bibnamefont
			{Tybell}}, \bibinfo {author} {\bibfnamefont {P.}~\bibnamefont {Paruch}},
		\bibinfo {author} {\bibfnamefont {T.}~\bibnamefont {Giamarchi}},\ and\
		\bibinfo {author} {\bibfnamefont {J.-M.}\ \bibnamefont {Triscone}},\
	}\bibfield  {title} {\bibinfo {title} {Domain wall creep in epitaxial
			ferroelectric
			$\mathrm{Pb}(\mathrm{Zr}_{\mathrm{0.2}}\mathrm{Ti}_{\mathrm{0.8}})\mathrm{O}_{\mathrm{3}}$
			thin films},\ }\href {https://doi.org/10.1103/PhysRevLett.89.097601}
	{\bibfield  {journal} {\bibinfo  {journal} {Phys. Rev. Lett.}\ }\textbf
		{\bibinfo {volume} {89}},\ \bibinfo {pages} {097601} (\bibinfo {year}
		{2002})}\BibitemShut {NoStop}%
	\bibitem [{\citenamefont {Casals}\ \emph {et~al.}(2021)\citenamefont {Casals},
		\citenamefont {Nataf},\ and\ \citenamefont {Salje}}]{CAS-21}%
	\BibitemOpen
	\bibfield  {author} {\bibinfo {author} {\bibfnamefont {B.}~\bibnamefont
			{Casals}}, \bibinfo {author} {\bibfnamefont {G.~F.}\ \bibnamefont {Nataf}},\
		and\ \bibinfo {author} {\bibfnamefont {E.~K.}\ \bibnamefont {Salje}},\
	}\bibfield  {title} {\bibinfo {title} {Avalanche criticality during
			ferroelectric/ferroelastic switching},\ }\href@noop {} {\bibfield  {journal}
		{\bibinfo  {journal} {Nature Communications}\ }\textbf {\bibinfo {volume}
			{12}},\ \bibinfo {pages} {345} (\bibinfo {year} {2021})}\BibitemShut
	{NoStop}%
	\bibitem [{\citenamefont {T{\"u}ckmantel}\ \emph {et~al.}(2021)\citenamefont
		{T{\"u}ckmantel}, \citenamefont {Gaponenko}, \citenamefont {Caballero},
		\citenamefont {Agar}, \citenamefont {Martin}, \citenamefont {Giamarchi},\
		and\ \citenamefont {Paruch}}]{TUC-21}%
	\BibitemOpen
	\bibfield  {author} {\bibinfo {author} {\bibfnamefont {P.}~\bibnamefont
			{T{\"u}ckmantel}}, \bibinfo {author} {\bibfnamefont {I.}~\bibnamefont
			{Gaponenko}}, \bibinfo {author} {\bibfnamefont {N.}~\bibnamefont
			{Caballero}}, \bibinfo {author} {\bibfnamefont {J.~C.}\ \bibnamefont {Agar}},
		\bibinfo {author} {\bibfnamefont {L.~W.}\ \bibnamefont {Martin}}, \bibinfo
		{author} {\bibfnamefont {T.}~\bibnamefont {Giamarchi}},\ and\ \bibinfo
		{author} {\bibfnamefont {P.}~\bibnamefont {Paruch}},\ }\bibfield  {title}
	{\bibinfo {title} {Local probe comparison of ferroelectric switching event
			statistics in the creep and depinning regimes in
			{P}b({Z}r$_{0.2}${T}i$_{0.8}$){O}$_3$ thin films},\ }\href
	{https://doi.org/10.1103/PhysRevLett.126.117601} {\bibfield  {journal}
		{\bibinfo  {journal} {Physical Review Letters}\ }\textbf {\bibinfo {volume}
			{126}},\ \bibinfo {pages} {117601} (\bibinfo {year} {2021})}\BibitemShut
	{NoStop}%
	\bibitem [{\citenamefont {Ponson}\ \emph {et~al.}(2006)\citenamefont {Ponson},
		\citenamefont {Bonamy},\ and\ \citenamefont {Bouchaud}}]{PON-06}%
	\BibitemOpen
	\bibfield  {author} {\bibinfo {author} {\bibfnamefont {L.}~\bibnamefont
			{Ponson}}, \bibinfo {author} {\bibfnamefont {D.}~\bibnamefont {Bonamy}},\
		and\ \bibinfo {author} {\bibfnamefont {E.}~\bibnamefont {Bouchaud}},\
	}\bibfield  {title} {\bibinfo {title} {Two-dimensional scaling properties of
			experimental fracture surfaces},\ }\href
	{https://doi.org/10.1103/PhysRevLett.96.035506} {\bibfield  {journal}
		{\bibinfo  {journal} {Phys. Rev. Lett.}\ }\textbf {\bibinfo {volume} {96}},\
		\bibinfo {pages} {035506} (\bibinfo {year} {2006})}\BibitemShut {NoStop}%
	\bibitem [{\citenamefont {Kuepferling}\ \emph {et~al.}(2023)\citenamefont
		{Kuepferling}, \citenamefont {Casiraghi}, \citenamefont {Soares},
		\citenamefont {Durin}, \citenamefont {Garcia-Sanchez}, \citenamefont {Chen},
		\citenamefont {Back}, \citenamefont {Marrows}, \citenamefont {Tacchi},\ and\
		\citenamefont {Carlotti}}]{KUE-23}%
	\BibitemOpen
	\bibfield  {author} {\bibinfo {author} {\bibfnamefont {M.}~\bibnamefont
			{Kuepferling}}, \bibinfo {author} {\bibfnamefont {A.}~\bibnamefont
			{Casiraghi}}, \bibinfo {author} {\bibfnamefont {G.}~\bibnamefont {Soares}},
		\bibinfo {author} {\bibfnamefont {G.}~\bibnamefont {Durin}}, \bibinfo
		{author} {\bibfnamefont {F.}~\bibnamefont {Garcia-Sanchez}}, \bibinfo
		{author} {\bibfnamefont {L.}~\bibnamefont {Chen}}, \bibinfo {author}
		{\bibfnamefont {C.}~\bibnamefont {Back}}, \bibinfo {author} {\bibfnamefont
			{C.}~\bibnamefont {Marrows}}, \bibinfo {author} {\bibfnamefont
			{S.}~\bibnamefont {Tacchi}},\ and\ \bibinfo {author} {\bibfnamefont
			{G.}~\bibnamefont {Carlotti}},\ }\bibfield  {title} {\bibinfo {title}
		{Measuring interfacial {Dzyaloshinskii-Moriya} interaction in ultrathin
			magnetic films},\ }\href {https://doi.org/10.1103/revmodphys.95.015003}
	{\bibfield  {journal} {\bibinfo  {journal} {Rev. Mod. Phys.}\ }\textbf
		{\bibinfo {volume} {95}},\ \bibinfo {pages} {015003} (\bibinfo {year}
		{2023})}\BibitemShut {NoStop}%
\end{thebibliography}
\end{document}

\onecolumngrid
\renewcommand{\thefigure}{S\arabic{figure}}
\renewcommand\theequation{S\arabic{equation}}
\renewcommand{\thesection}{S\arabic{section}}
\renewcommand{\thetable}{S\arabic{table}}
\setcounter{figure}{0}
\setcounter{equation}{0}

\section*{SUPPLEMENTARY INFORMATION}
\setcounter{equation}{0}
\setcounter{figure}{0}
\subsection*{S1: Sample preparation and Experimental details}
The sample investigated consists of Ta(5nm)/Co$_{20}$Fe$_{60}$B$_{20}$(1nm)/MgO(2nm)/Ta(3nm) thin film, grown by sputtering on Si(001)/SiO$_2$ at room temperature using a \textit{Singulus Timaris} deposition system. The as-grown sample was then annealed at 300 $^\circ$C for two hours to promote CoFeB crystallization into the bcc structure.  

The MOKE microscope consists of a home-modified Zeiss Axioskop 2 Plus, equipped with a mercury arc lamp, a $20\times$ objective lens (numerical aperture = 0.5) and a Thorlab CCD camera. Images were acquired in a polar configuration with a camera exposure time of 200 ms and a frame acquisition rate of 5 fps. Magnetic bubbles were nucleated by applying a perpendicular field pulse of amplitude 9.5 mT and duration  50 $\mu$s through a small coil. The bubbles were then grown under a continuous perpendicular field (0.13 mT -- 0.16 mT) applied through a perpendicular electromagnet.  To improve the statistics and make the data analysis more robust, measurements are repeated at least 16 times for each value of the applied field. The bubble domain nucleates reproducibly at the same location, probably corresponding to the site of an extrinsic defect where the perpendicular anisotropy constant is locally lowered.
\subsection*{S2: Resolving the wall dynamics}
\label{sec:supI}
\begin{figure}
\includegraphics[width=1\textwidth]{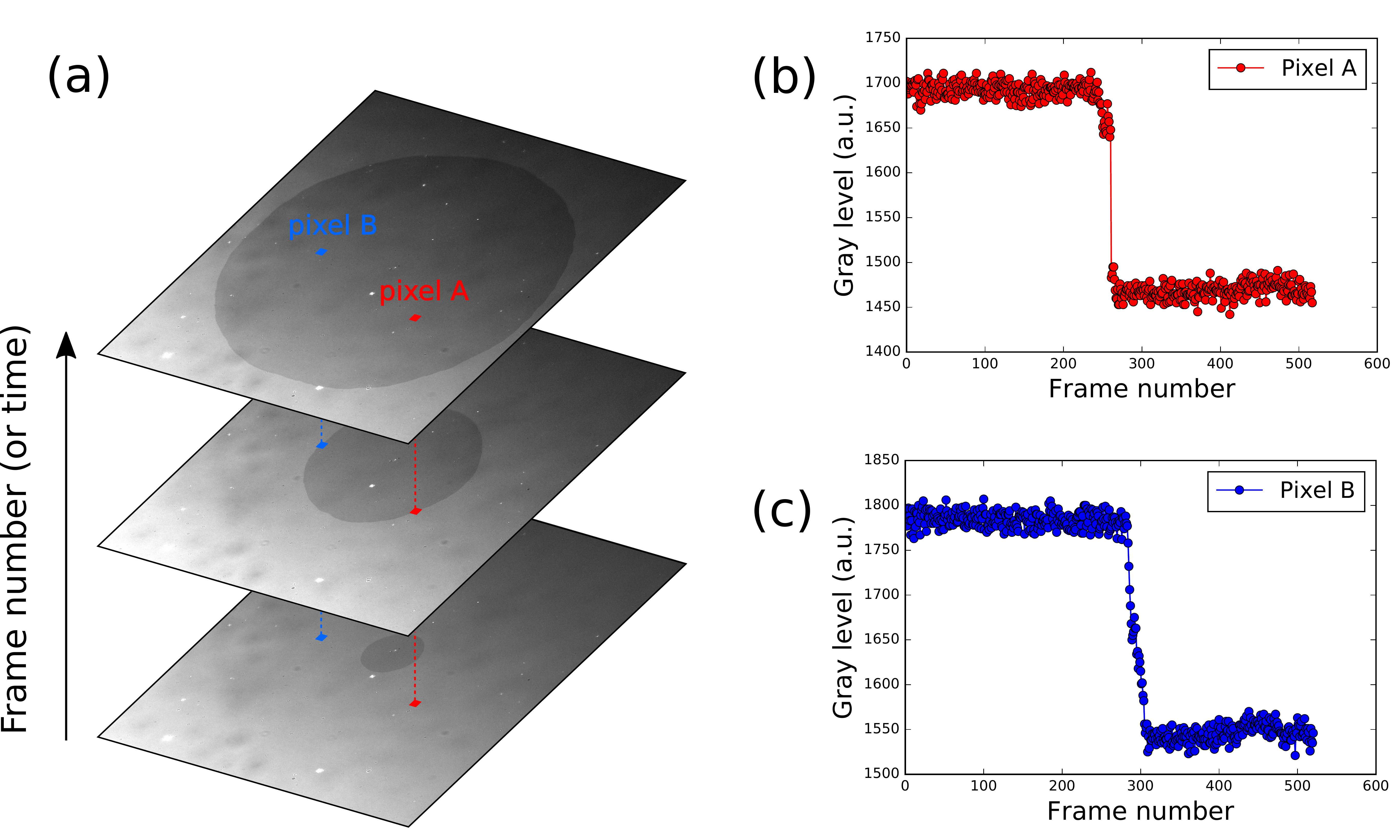}
\caption{\textbf{Schematic illustration of the procedure used to determine the position of the domain wall.} (a) All the raw images collected within one measurement of bubble expansion are stacked together into a 3D array. Marked in red and blue are two representative pixels whose gray level profiles are shown in (b) and (c), respectively. While pixel A switches its gray level abruptly, the transition occurs more slowly for pixel B. High/low gray level values correspond to bright/dark MOKE contrast, respectively.}
\label{fig:SI1}
\end{figure}

The procedure implemented to determine the position of the bubble domain wall during its field-driven motion relies on the detection of the time at which each pixel changes its gray level. This method requires that all the images captured every 200 ms during the growth of the bubble are stacked together into a 3D array, as schematically shown in Fig.~\ref{fig:SI1}(a). The bubble boundary is then identified by looking ``vertically'' across the stack, i.e. by plotting the gray level variation for each individual pixel as a function of the frame number within the stack. If the domain wall does not pass through a pixel, then the gray level of that pixel will remain approximately constant throughout the stack, being either bright or dark, depending on the magnetization direction at the pixel location. On the other hand, a rapid change of gray level in a pixel indicates a switch in magnetization direction or, in other words, a passing domain wall (see Fig.~\ref{fig:SI1}(b) and (c)). This pixel can then be considered as being part of the domain wall at the time at which the switching is observed. By analysing each individual pixel in such a way, the position of the domain wall is eventually detected with very good accuracy for each frame of the stack. Indeed, this method is found to provide more reliable and robust results than what would otherwise be possible by applying commonly used edge detection algorithms to the original ``horizontal'' 2D images.   

In ideal conditions, each pixel should switch its gray level abruptly, within one frame, as is the case shown in Fig.\ref{fig:SI1}(b). In reality, non-optimal focusing conditions, together with light inhomogeneities and drift introduce a degree of noise that can broaden the transition, resulting in gray level profiles as shown in Fig.\ref{fig:SI1}(c). In any case, the switching is identified with the frame in which the biggest gray level variation occurs. This is mathematically achieved by performing a cross-correlation between the gray level profile and a step function. Denoting the gray-level vs frame-number function as $f$ and the step function as $g$, the cross-correlation between the two is defined as:

\begin{equation}
(f \star g)[n] = \sum_{m} f(m)g(m + n),
\label{eq:cross_corr1}
\end{equation}

where $m$ is the frame number and $n$ is an integer running along the frame number axis. With the following choice for the step function,
    
\begin{equation} 
g(m) = 
\left \{ 
	\begin{tabular}{cc}
	$+1$ & \hspace{3mm} for $m < m_s$\\	
	$-1$ & \hspace{3mm} for $m > m_s$\\
		\end{tabular}
\right.,
\label{eq:step_fun}
\end{equation}

the cross-correlation becomes:
\begin{equation}
(f \star g)[n] = \sum_{m < m_s} f(m)~-\sum_{m > m_s} f(m),
\label{eq:cross_corr2}
\end{equation}

where $m_s$ runs again along the frame number axis. Eq.~\ref{eq:cross_corr2} is calculated for each $m_s$, and the value of $m_s$ for which the cross-correlation is found to be maximum provides the frame number at which the switching takes place.  

\subsection*{S3: Clustering algorithm}\label{sec:supII}
In this section, we explain  the algorithm used for clustering frame events. 

We consider the $i$-th frame event occurring at time $t_i$. It  activates $S_i$  pixels at coordinates $\{ x_{i, n}, y_{i, n} \}_{n=1}^{S_i}$. We want to find all the subsequent frame events that are correlated with $i$.
\begin{itemize}
    \item Consider all the events $j$ for which $t_j > t_i$ and $t_j-t_i \leq \Delta t$.
    \item the event $j$ is connected to $i$ if their minimal distance  is less than $\Delta s$, namely if:
    \begin{equation}
        \min_{(n,m)} |x_{i,n}-x_{j,m}| + |y_{i,n} - y_{j,m}| \leq \Delta s
    \end{equation}
\end{itemize}
Repeating the above procedure we can construct an adjacency matrix $A_{ij}$ for which $A_{ij}=1$ if $j$ is connected to $i$. Each connected component of the graph $A_{ij}$ corresponds to an individual cluster. 
We set $\Delta s =2$ because the pixels are organized on a square grid.

\begin{figure}[H]
\centering
\includegraphics[width=.5\textwidth]{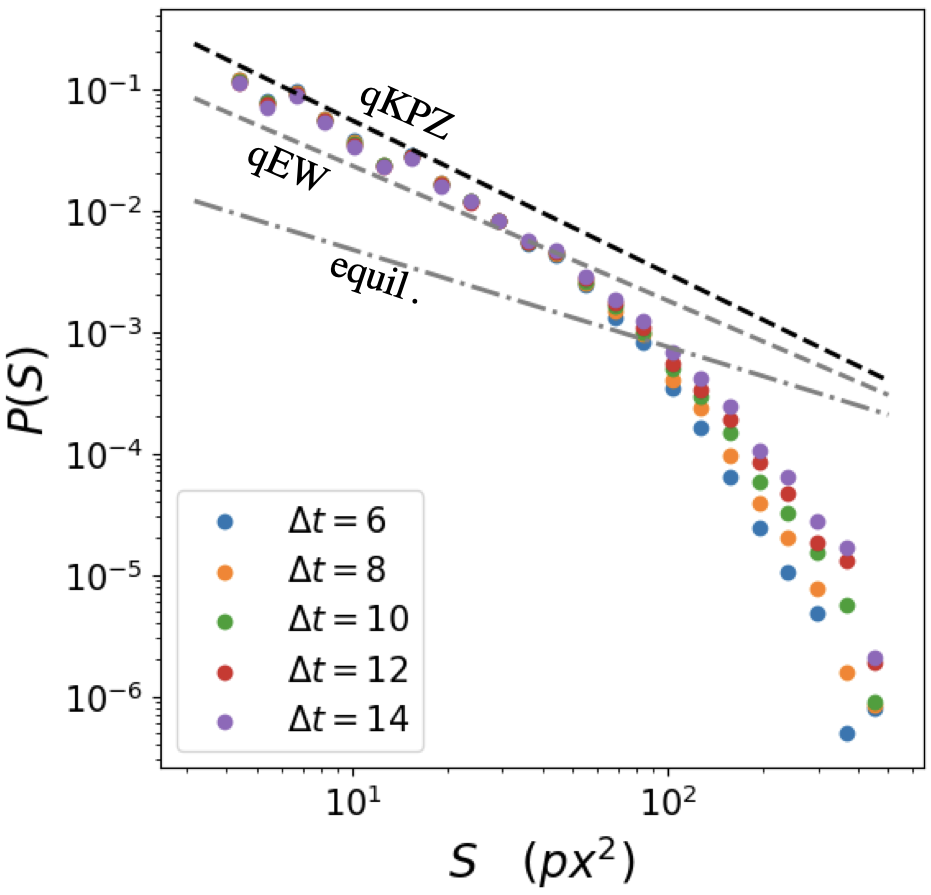}
\caption{ Cluster size distribution $P(S)$ obtained for $\Delta s = 2$, field $H=0.14$ mT and different $\Delta t$. }
\label{fig:comp_cluster}
\end{figure}

Using different $\Delta t$ does not affect our results sensibly as visible in Fig.~\ref{fig:comp_cluster}. Indeed for the cluster size distribution, the effect of a larger $\Delta t$ is to increase the cutoff without affecting the power-law exponent.

 \subsection*{S4: Longitudinal length of the cluster}
To extract the longitudinal length of the cluster, we first need to determine its local growth direction. This is non-trivial because the geometry of the domain wall is circular and not directed as for the numerical interfaces in \cite{FER-17}. To achieve this task, we first center the pixels belonging to a cluster such that its set of coordinates $\{x_n, y_n\}_{n=1}^S$ has zero mean. Then we construct the covariance matrix:
\begin{equation} 
C_{\rm cluster} = \begin{pmatrix}
\frac{1}{S} \sum_n x_n^2 & \frac{1}{S} \sum_n x_n y_n \\
\frac{1}{S} \sum_n x_n y_n & \frac{1}{S} \sum_n y_n^2 
\end{pmatrix}
\end{equation}
 The normalized eigenvector $\vec{v}_+$ associated with the largest eigenvalue $\lambda_+$ of $C_{\rm cluster}$ gives the (local) longitudinal direction of the cluster, the eigenvector associated with the other eigenvalue $\lambda_-$ gives the growth direction. 
Moreover, $\sqrt{\lambda_+}$ and $\sqrt{\lambda_-}$ quantify the spread of the cluster points along the two directions and thus  $\sqrt{\lambda_+ \lambda_-}$ is proportional to the cluster size. Using  $S \sim \pi a^2 \sqrt{\lambda_+ \lambda_-}$ to determine the constant $a$, we associate to each cluster a longitudinal length  $\ell = a \sqrt{\lambda_+}$. 
\bibliography{creep,dws}